\documentclass{article}

\usepackage{verbatim} 
\usepackage{float}

\usepackage{PRIMEarxiv}

\usepackage[utf8]{inputenc} 
\usepackage[T1]{fontenc}    
\usepackage{hyperref}       
\usepackage{url}            
\usepackage{booktabs}       
\usepackage{amsfonts}       
\usepackage{nicefrac}       
\usepackage{microtype}      
\usepackage{lipsum}
\usepackage{fancyhdr}       
\usepackage{graphicx}       

\usepackage{mathtools}
\usepackage{amsthm,amsmath,amssymb,amsfonts,exscale,latexsym,float,eucal}

\usepackage{lineno}
\usepackage{soul}
\usepackage{color, xcolor}
\usepackage{mathrsfs,amsmath}
\usepackage[linesnumbered,ruled]{algorithm2e}
\usepackage{cite}
\usepackage{hyperref}
\usepackage{cleveref}
\hypersetup{colorlinks = true,
	citecolor = blue}

\title{Refractive index tomography with a physics based optical neural network}

\author{
  Delong Yang \\
  School of Optics and Photonics, Beijing Institute of Technology \\
  Beijing, China \\
   \And
  Shaohui Zhang*\\
  School of Optics and Photonics, Beijing Institute of Technology \\
  Beijing, China \\
  \texttt{*zhangshaohui@bit.edu.cn} \\
   \And
  Yao Hu \\
  School of Optics and Photonics, Beijing Institute of Technology \\
   \And
  Qun Hao*\\
  School of Optics and Photonics, Beijing Institute of Technology \\
  \texttt{*qhao@bit.edu.cn}
}

\begin{document}
\maketitle

\begin{abstract} 
The non-interference three-dimensional refractive index(RI) tomography has attracted extensive attention in the life science field for its simple system implementation and robust imaging performance. However, the complexity inherent in the physical propagation process poses significant challenges when the sample under study deviates from the weak scattering approximation. Such conditions complicate the task of achieving global optimization with conventional algorithms, rendering the reconstruction process both time-consuming and potentially ineffective. To address such limitations, this paper proposes an untrained multi-slice neural network(MSNN) with an optical structure, in which each layer has a clear corresponding physical meaning according to the beam propagation model. The network does not require pre-training and performs good generalization and can be recovered through the optimization of a set of intensity images. Concurrently, MSNN can calibrate the intensity of different illumination by learnable parameters, and the multiple backscattering effects have also been taken into consideration by integrating a "scattering attenuation layer" between adjacent "refractive index" layers in the MSNN. Both simulations and experiments have been conducted carefully to demonstrate the effectiveness and feasibility of the proposed method. Experimental results reveal that MSNN can enhance clarity with increased efficiency in RI tomography. The implementation of MSNN introduces a novel paradigm for RI tomography.
\end{abstract}

\section{Introduction}
Remarkable three-dimensional(3D) visualization of biological processes is of paramount importance for biological research. This is particularly challenging due to the transparent or semi-transparent characteristics of numerous biological cells. Direct microscopic observation often results in a low imaging contrast, missing much information. For improving the imaging contrast and signal-to-noise ratio(SNR) of 3D distributed bio-samples, fluorescence microscopy is one of the most widespread solutions\textsuperscript{\cite{ross2000systematic,rustom2004nanotubular,okuda2002mitochondrial,rizzuto1995chimeric}}, and has inspired several representative super-resolution methods, such as Super-resolution microscopy(STED)\textsuperscript{\cite{betzig2006imaging,hell1994breaking}}, Stochastic optical reconstruction microscopy(STORM)\textsuperscript{\cite{rust2006sub,bates2007multicolor}}, Photoactivation localization microscopy(PALM)\textsuperscript{\cite{manley2008high}}, et al. Combined with necessary mechanical position scanning techniques, confocal designs\textsuperscript{\cite{paddock1999confocal}}, multi-photon microscopy\textsuperscript{\cite{zipfel2003nonlinear}} and light sheet microscopy\textsuperscript{\cite{keller2008reconstruction}} can be implemented, fluorescence microscopy enables the achievement of high-resolution 3D imaging of typical biological samples, with both high signal-to-noise ratio(SNR) and spatial resolution. Nevertheless, fluorescence microscopy necessitates exogenous labeling. And since some cellular characteristics or structures may not be detectable using fluorescence, it is unable to probe the total internal structure of biological samples.

Unlike fluorescent labeling methods, phase imaging by measuring the change in the optical path difference of light is another approach that can effectively observe the structure information of transparent and semi-transparent samples. Combining with angularly-resolved measurements and tomography reconstruction algorithms, 3D refractive index(RI) imaging can be achieved, termed optical projection tomography\textsuperscript{\cite{sharpe2002optical, alanentalo2007tomographic, correia2015accelerated}}. Although optical tomography can produce high-quality 3D images, the interferometry-based method for obtaining phase distribution under each angular projection is highly sensitive to environmental conditions and prone to speckle noise. Therefore the corresponding system construction cost is high, and the system is less robust.

Besides interferometry, phase retrieval with intensity images utilizing optimization algorithms is also a widely used phase imaging technique. Such methods are not affected by laser speckle noise and are robust to environmental disturbances, so the implementation and operation costs of the systems are also relatively low. Nevertheless, phase imaging from intensity images is not trouble-free. The ability of an optimization algorithm to find a satisfactory global solution depends on the inverse problem's complexity, the physical constraints used, and the chosen initial value. RI tomography from intensity images, referred to as intensity diffractive tomography(IDT)\textsuperscript{\cite{tian20153d, horstmeyer2016diffraction, ling2018high, li2019high, zuo2020wide, zhou2022accelerated, li2022transport}}, is a complex inverse problem involving a large number of unknown parameters and complex multiple scattering physical processes. Conventional IDT approaches, whether they utilize the multi-slice model in the space domain or Ewald’s diffractive sphere model in the Fourier domain, have the problems of being seriously time-consuming, slow convergence, and easy to fall into local minima traps\textsuperscript{\cite{YangZewen2022Pro}}. 

In recent years, many researchers have been actively exploring appropriate deep-learning methods to achieve better performance in computational imaging including tomography. While deep learning has been demonstrated to improve the speed and quality of tomography across various modalities\textsuperscript{\cite{kang2017deep, jin2017deep, thanh2018deep, aggarwal2018modl, sun2018efficient, li2018deep}}, conventional deep learning approaches remain unsuitable for certain measurement applications, especially RI tomography of biological samples. Since the conventional deep learning methods are mostly data-driven schemes, the inferential capabilities of the convolution neural networks(CNNs) are from the "experience" gleaned from huge datasets\textsuperscript{\cite{lecun2015deep, xing2017deep, moen2019deep}}. However, constructing a comprehensive dataset for biological tomography remains a significant challenge. Although a physics-based simulator has been proposed to generate training datasets for CNNs in biological tomography, the results are still limited by the mismatch between simulation and experimental data.\textsuperscript{\cite{matlock2021physical}}. Moreover, the accuracy of results from conventional CNNs, when applied to completely novel samples, remains questionable.\textsuperscript{\cite{barbastathis2019use, zuochao2022deep}}.

To address the aforementioned challenges, researchers have proposed the use of neural networks, integrated with physical models, to enhance the accuracy of imaging. In 2020, Horstmeyer Roarke proposed the deep prior diffraction tomography\textsuperscript{\cite{zhou2020diffraction}} which employs neural networks as the phase retrieval algorithm and uses a light scattering model as the physical verification to improve the authenticity of the imaging. Guohai Situ practiced a similar scheme in coherent diffraction imaging\textsuperscript{\cite{wang2020phase}} in the same year. In 2022, Ulugbek S. Kamilov proposed the deep continuous artefact-free RI field, which uses the neural field network to implement phase retrieval. The sparse representation capability of neural field networks significantly reduces memory usage and addresses the issue of missing cones in intensity diffraction tomography\textsuperscript{\cite{liu2022recovery}}. However, these physics-based methods still employ conventional neural network structures as inverse operation networks. The optimization process essentially involves searching for the optimal inverse problem model within the vast parameter space of CNNs, a process that remains time-consuming.

In this work, we remodel the structure of conventional neural networks\textsuperscript{\cite{lin2018all}} to incorporate both the beam propagation model and the scattering model, as opposed to solely relying on conventional architectures. The beam propagation method\textsuperscript{\cite{maiden2012ptychographic, chowdhury2019high, hu2022multi}}, dividing the RI of the sample into layers, is adapted within the structure of our neural network\textsuperscript{\cite{kamilov2015learning, jiang2018solving, yang2022fourier}}. The beam propagation model imposes constraints on the backward gradient and helps verify the authenticity of the imaging results. The neural network we propose, which incorporates an optical structure, is designated as the multi-slice neural network(MSNN). To enhance imaging quality, we incorporate dark field raw data into the optimization\textsuperscript{\cite{chang2020computational}}. We extend the MSNN connection for learnable parameters\textsuperscript{\cite{yanny2022deep}} to calibrate the intensity between different LEDs, reducing the requirements for acquisition hardware and the need for image preprocessing. The exceptional fitting capability of MSNN even offers the potential to take the higher-order scattering within the sample into phase retrieval. We take into account the backscattering field as predicted by the scattering model. The multiple backscattering effects have also been taken into consideration by integrating a "scattering attenuation layer" between adjacent "refractive index" layers in the MSNN. This approach ensures that our network architecture more accurately mirrors the actual process of light field propagation within the sample. We demonstrate that the implementation of adaptive recovery of backscatter intensity using MSNN during C. elegans imaging can enhance the quality of the images.

\section{method}

\subsection{The principle of the multi-slice neural network(MSNN)}
Multi-slice beam propagation method separates 3D objects into a series of thin layers, where the light wave through the sample is modeled by sequential layer-to-layer propagation of the light field. The structure of the beam propagation model is very similar to the hierarchical structure of the Deep Neural Networks(DNNs). And there are many optimizers\textsuperscript{\cite{ruder2016overview}} for DNNs to quickly converge to the global optimum. In MSNN, We model the $m$th layer of the 3D object as $L_{m}(r) = n(r,m\Delta z)-n_{media}$. Mathematically, the diffraction propagation in MSNN can be recursively written as:

\begin{gather}
P(r,\Delta z) =  exp(j2\pi\Delta z(\dfrac{n_{media}}{\lambda}^2-\Vert u\Vert^2)^{1/2}) \\
t_{m}(r,\Delta z) = exp(j2\pi\Delta z \dfrac{n(r,m\Delta z)-n_{media}}{\lambda}) \\
U_{m+1}(r) = t_{m}(r,\Delta z) \cdot \mathscr{F}^{-1}\{{P(r,\Delta z) \cdot \mathscr{F}\{U_{m}(r)\}}\}
\label{Eq.(3)}
\end{gather}
where $P(r,\Delta z)$ denotes the angular spectrum diffraction equation that propagates a light field by distance $\Delta z$, $r$ denotes the 2D spatial position vector, $u$ denotes the 2D spatial frequency space coordinates vector, $t_{m}(r,\Delta z)$ denotes the phase modulation by the $m$th layer, $U_{m}$ and $U_{m+1}$ are the input and output light field of $m$th layer in MSNN.
The boundary condition to initialize the recursively Eq.(\ref{Eq.(3)}) is the incident plane wave 
illuminating the sample, $U_{0}(r)=exp(jk_{illu}r)$ where $k_{illu}$ is the illumination wave vector at a particular angle. The experiment system and the principle of MSNN are shown in Fig.\ref{Fig.1}.

\begin{figure}
	\centering
	\includegraphics[scale=0.23]{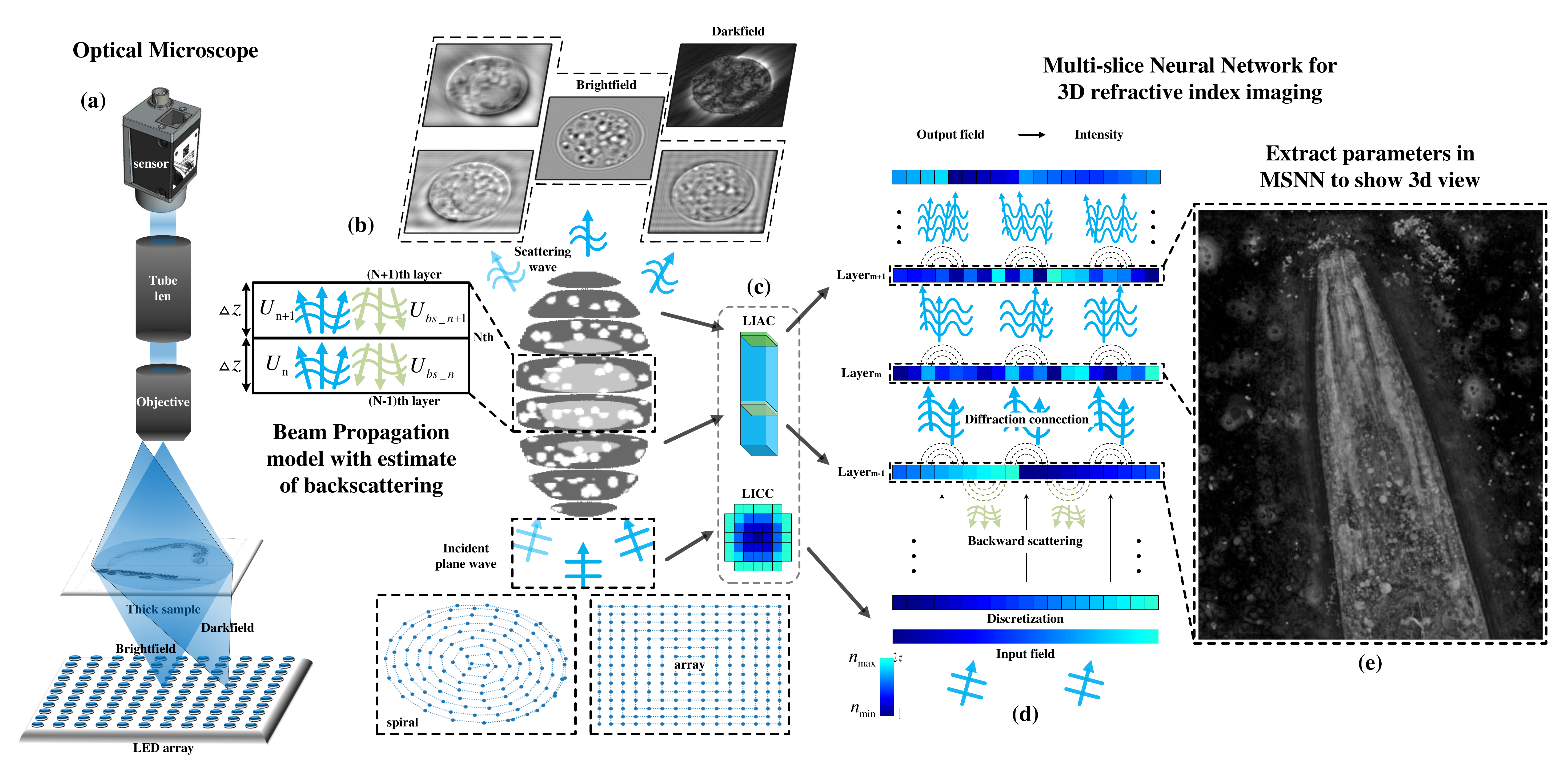}
	\caption{The principle of MSNN. (a)The optical system of the RI tomography. (b)The beam propogation model with estimate of backscattering. The lighting mode can be selected array lighting or spiral lighting. The bright-field and dark-field images are simulation captured images of a cell phantom.(c)The backscattering of different layers in the sample is predicted in MSNN with Learnable Intensity Attenuation Coefficient(LIAC). The inconsistency in intensity of various incident plane waves is corrected by Learnable Intensity Compensation Coefficient(LICC) in MSNN. (d)The structure of multi-slice neural network(MSNN). After the optimization, we will extract the internal parameters of the neural network as the RI reconstruction results of the sample. (e)The reconstruction result of the head of a C.elegan. The image is a two-dimensional projection of the reconstruction result on the x-y plane.}
	\label{Fig.1}
\end{figure}
Simultaneously, based on the 1st Born approximation, the scattering model that suggests the light field through the 3D sample including the incident field and scattered field stimulated by the 3D scattering potential $V(r,z) = k^2_{0}(n^2_{media}-n^2(r,z))$. The total light field through the 3D sample can be expressed as:

\begin{gather}
U_{total}(r, z) = U_{in}(r, z) + U^{(born)}_{s}(r, z) \\
U_{in}(r, z) = \mathscr{F}^{-1}\{P(r,\Delta z)\cdot \mathscr{F}\{U_{in}(r, z-\Delta z)\}\} \\
U^{(born)}_{s}(r, z) = \iiint G(r-r',\Delta z)U_{in}(r',z-\Delta z)V(r',z)d^3r'
\label{Eq.(6)}
\end{gather}
$U^{(born)}_{s}(r, z)$ denotes the scattered field under the 1st Born approximation. The beam propagation model and the 1st Born approximation scattering model both consider the propagation of the input field according to the angular spectrum diffraction method, but they are different in the prediction of the scattering field. And the previous work has verified that both models are effective for 3D imaging of weakly scattered thick samples\textsuperscript{\cite{chowdhury2019high,chen2020multi}}. In scattering model, Eq.(\ref{Eq.(6)}) suggests that the superposition field of spherical waves computed by the Green function is actually the total scattering field including the forward scattering field and backscattering field, and we proved that when the layer is thin enough, the forward field and backscattering field can be approximate have same distribution in phase but different in intensity(Supplementary 2). We transferred the conclusion to the beam propagation model and got better imaging results in the experiments. So we still employ the beam propagation model to predict the backscattering field and attenuate the incident field in the layer of the forward propagation in MSNN, we set learnable parameters $\omega_{m}$ to predict the intensity of backscattering field to attenuate forward field, the parameters $\omega_{m}$ which we called Learnable Intensity Attenuation Coefficient(LIAC) are optimized by MSNN. The forward propagation in MSNN can be expressed as:

\begin{gather}
U_{m+1}(r) = t_{m}(r,\Delta z) \cdot \mathscr{F}^{-1}\{{P(r,\Delta z) \cdot \mathscr{F}\{U_{m}\}}\} / \omega_m
\end{gather}
The exit electric-field, $U_{M}(r)$, accounts for the light field passing through the sample. When the imaging system is focus at the center of the sample, we need to refocus the exit light field $U_{M}(r)$ to the center of the sample, where $M$ denotes the total layers of the model. The final light field and intensity distributions at the image plane are:

\begin{gather}
U_{predict}(r) = U_{M}(r) \mathscr{F}^{-1}\{C(u, k_0)\cdot{P(r,-\dfrac{M\Delta z}{2}) \cdot \mathscr{F}\{U_{M}\}}\} \\
I_{predict}(r) = |U_{predict}(r)|^2
\end{gather}
where $C(u, k_0)$ denotes the coherent transfer function of the system. The light field captured by the objective $U_{predict}(r)$ accounts for the accumulation of the diffraction and multiple-scattering processes(including the backscattering) that occurred during optical propagation through the ($-\dfrac{M\Delta z}{2}$,$\dfrac{M\Delta z}{2}$) around the focal plane.

\subsection{Adaptive intensity calibration between different input fields in MSNN}
In practical experiments, to enhance lateral and axial resolution and improve the success rate of 3D imaging, we need to provide plane waves $k_{i}$ from various angles. When $k_{i} > NA_{obj}$, the camera captures dark-field images which require longer exposure times to enhance the signal-to-noise ratio. During the image capture process, we set distinct exposure times for different illumination angles. However, the relationship between exposure time length and sensor responsiveness is not strictly linear. Despite raw data preprocessing according to exposure times, this discrepancy could potentially introduce complications in solving the inverse problem.

\begin{algorithm}[H]
	\DontPrintSemicolon
	\textbf{Input:}The illuminate $k_{illu}$ of $N$ LEDs, $M$ layers of imaging axial size $M\Delta z$.\\
	\textbf{Data:}Measured intensities $\{I^n_{measure}(r)\}_{n=1}^{N}$.\\
	\textbf{Hyperparameters:}The parameters of the microscope system, the step size of optimization $\alpha$, the TV regularization coefficient $\beta$, the learning rate of $\mathbf{LIAC}$ $\omega_m$ and $\mathbf{LICC}$ $\gamma_n$, max number of iteration $I$\\
	\textbf{Initialization:}The RI in the layer of MSNN $\{L_m(r)\}_{m=1}^M=0$.\\
	\textbf{Return:}3D RI of the sample. \\
	\For{$i = 1:I$}{
		\For{$n=1:N$}{
			$k_{illu} = k_n$, $I_{gt}(r) = I^n_{measure}(r)$ \\
			$U_{0}(r)=\gamma_n \cdot exp(jk_{illu}r)$ \\
			\For{$m=1:M$}{$L_m(r) \gets$ mth layer of MSNN \\
				$t_{m}(r,\Delta z) \gets exp(j2\pi\Delta z \dfrac{L_m(r)}{\lambda})$ \\
				$U_{m+1}(r) \gets t_{m}(r,\Delta z) \cdot \mathscr{F}^{-1}\{{P(r,\Delta z) \cdot \mathscr{F}\{U_{m}(r)\}}\} / \omega_m$}
			$I_{predict}(r) \gets |U_{M}(r) \mathscr{F}^{-1}\{C(u, k_0)\cdot{P(r,-\dfrac{M\Delta z}{2}) \cdot \mathscr{F}\{U_{M}\}}\}|$ \\
			$Loss \gets L_1(I_{predict}(r), I_{gt}(r))$ \\
			$Loss.autograd().backward()$ by optimizer $Adam(\alpha)$}
		$RI \gets $ layers of MSNN \\
		$RI_{reg} \gets TV_{3D}(RI,\beta)$ \\
		MSNN $\gets RI_{reg}$}
	\caption{3D Intensity-based RI imaging with MSNN}
\end{algorithm}

Furthermore, due to limitations in the accuracy of the experimental system, the intensity of the high-order diffracted input light field(provides illumination for darkfield) - which is used as an initial condition of MSNN - may be inaccurate. To address discrepancies in the response gray value in raw data caused by varying sensor exposure time settings at different illumination angles, and to compensate for the intensity of higher-order diffracted light, we introduce learnable parameters $\gamma_{n}$ for each illumination angle. Consequently, the incident wave can be expressed as follows:

\begin{gather}
U^n_0(r) = \gamma_n \cdot exp(j k_{illu} r)
\end{gather}
We define $\gamma_n$ as the Learnable Intensity Compensation Coefficient(LICC) for the nth incident plane wave. In our model, LEDs in different positions each have independent coefficients. These coefficients can be learned and optimized within the Multi-Slice Neural Network(MSNN) during imaging. Consequently, the MSNN identifies the optimal set of LICCs $(\gamma_1...,\gamma_n...,\gamma_N)$ for the LEDs, thereby enhancing the 3D imaging process.

\begin{figure}
	\centering
	\includegraphics[scale=1]{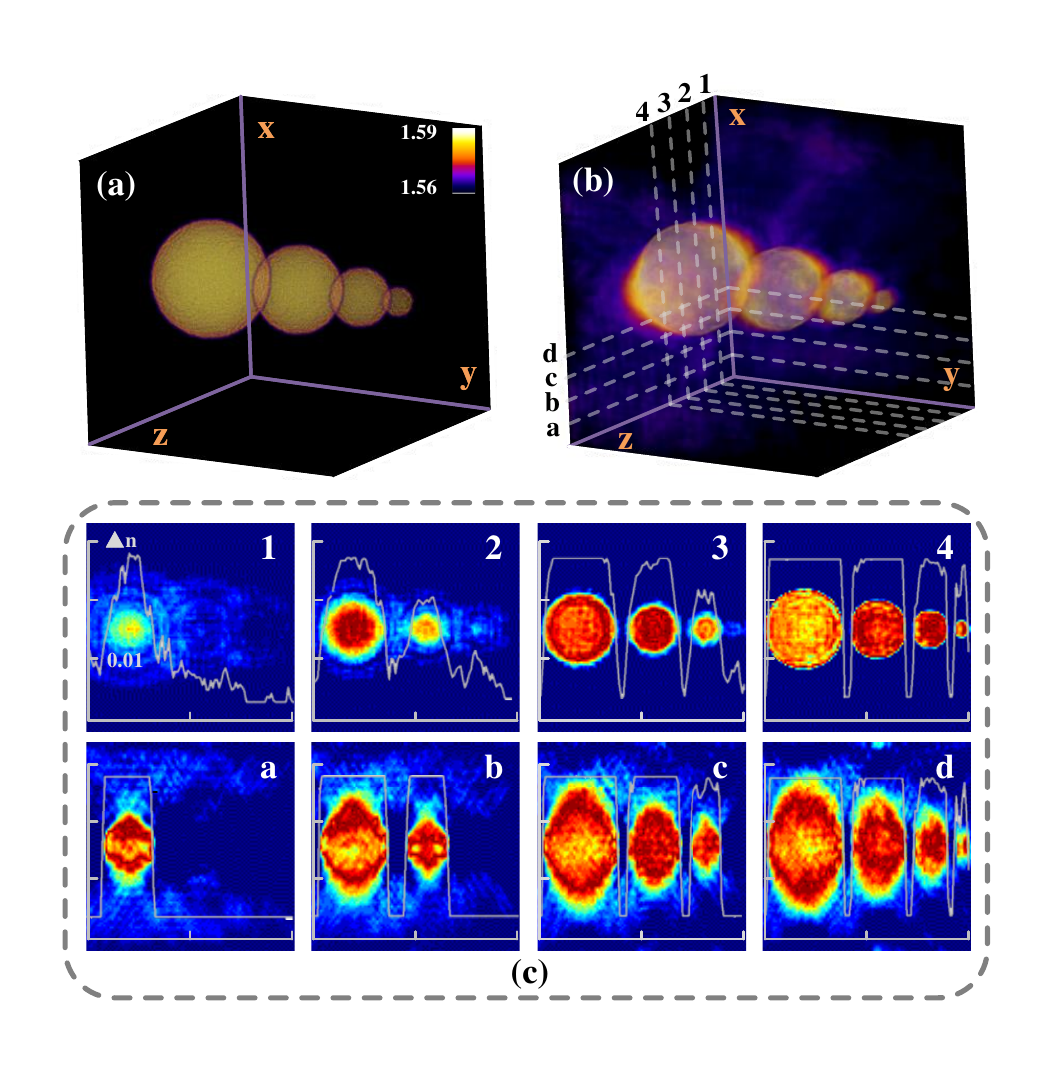}
	\caption{The reconstruction results for 4 microspheres of varying sizes (a)This panel depicts the 3D ground truth of the microspheres, with radius of 4um, 3um, 2um and 1um. (b)The 3D reconstuction results of the microspheres. (c)The reconstructions at depths of 1um, 2um, 3um and 4um in both the x-y and y-z planes.}
	\label{Fig.2}
\end{figure}
\subsection{RI tomography with MSNN}
MSNN is employed to solve the following optimization with an objective consisting of a measurement loss $\mathcal{L}$ and regularizer $\mathcal{R}$:

\begin{gather}
argmin\{\mathcal{L}(MSNN(k_illu), I_{gt}(r)) + \beta\cdot \mathcal{R}(MSNN)\}
\end{gather}
where $\mathcal{L}$ denotes the L1 loss. $\mathcal{R}$ denotes the regularization term. We use the 3D total variation(TV) norm as the regularization loss within the MSNN. The parameter $\beta$ is manually adjusted to optimize the strength of the regularization, with a larger punishment in the axial direction typically yielding better results. For the MSNN updates, we employ the Adam optimizer\textsuperscript{\cite{kingma2014adam}}, which adaptively adjusts the learning rate based on raw data. This process for 3D Intensity-based RI imaging with MSNN is summarized in Algorithm 1.

\section{Simulations}
To assess its capability for quantitative 3D RI reconstruction, we initially create a pure phase phantom composed of microspheres of varying radius. These microspheres have a RI of 1.59, while the medium has an RI of 1.56. The ideal microspheres, with diameters ranging from 1 to 4 µm and an RI of 1.59, are immersed in a medium matching the RI of 1.56. The numerical aperture(NA) of both the objective(Magnification 40X) and the illumination in the simulation is 0.7. We sequentially illuminate angle-varied plane waves, with a center wavelength of 532 nm, through the sample and generate 225 captured intensity images. Each image contains 100 × 100 pixels, with a sensor pixel size of 4µm. Figure 2 shows the 3D imaging results of the microspheres, with MSNN producing a satisfactory quantitative 3D RI. However, larger radius microspheres demonstrate larger axial artifacts, attributable to the problem of low-frequency missing cones in the frequency domain(detailed in Supplementary 1).

Subsequently, we create a synthetic cell phantom containing intricate details for 3D RI recovery, with an RI that ranges from 1.33 to 1.38. Figure 3 depicts the imaging results of the cell phantom by MSNN, using 225 measurements. Even though it is subjected to the missing cone problem and compromises axial resolution for low spatial frequencies, MSNN manages to reconstruct a satisfactory quantitative 3D RI. Simulation was performed with PyTorch running on a desktop computer equipped with Intel(R) Core(TM) i5-10500 CPU at 4GHz 32GB RAM CPU and NVIDIA's GeForce GTX 3090 GPU. The total imaging time to complete 10 iterations for a volume of 100$\times$100$\times$100 voxels was 40 seconds.

\section{Quantitative verification}
\subsection{System setup}
To demonstrate the improvement of MSNN, we use a two-slice test sample consisting of two resolution targets, one placed above the focal plane and the other axially spaced apart and rotated relative to the first one. We build a microscope system as shown in Fig.\ref{Fig.1} with a 4x objective(NA$\approx$0.1). We use a LED array(15$\times$15) for varied-angle illuminations. The LED array is located 100mm from the sample plane and emits a light with a wavelength of 473nm and a bandwidth of 20nm. The LED array is controlled by an stm32 microcontroller and is synchronized with the camera to scan through the LEDs at camera-limited speeds. Our camera is capable of 50 frames per second at full frame(2448$\times$2048 pixels, pixel size 3.45um) and with 8-bit data. However, for dark-field images, we opt for longer exposure times. The raw data, constrained to a small 360x360 pixel resolution region, is used to recover a high-resolution complex field with a 4X increase in pixel quantity. The computing platform is aligned with the simulation. The total imaging time to complete 50 iterations of a 720$\times$720$\times$2 voxels volume was 1 minutes.

\begin{figure}
	\centering
	\includegraphics[scale=0.5]{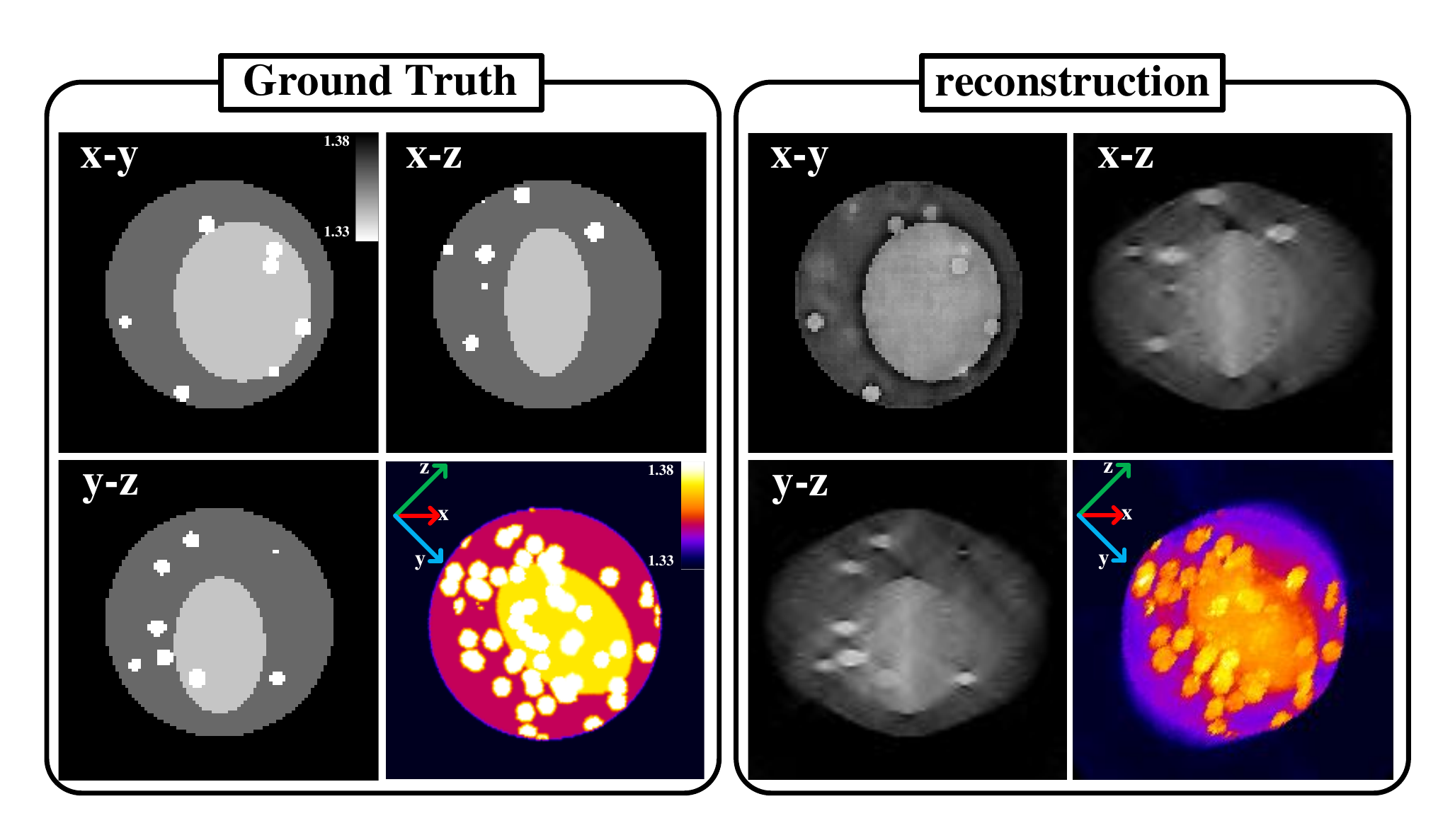}
	\caption{The middle slice of x-y,x-z,y-z plane of the ground truth and imaging with MSNN. And the 3D projection views in a tilt angle are present.}
	\label{Fig.3}
\end{figure}

\subsection{Experiment results}
\begin{figure}
	\centering
	\includegraphics[scale=0.16]{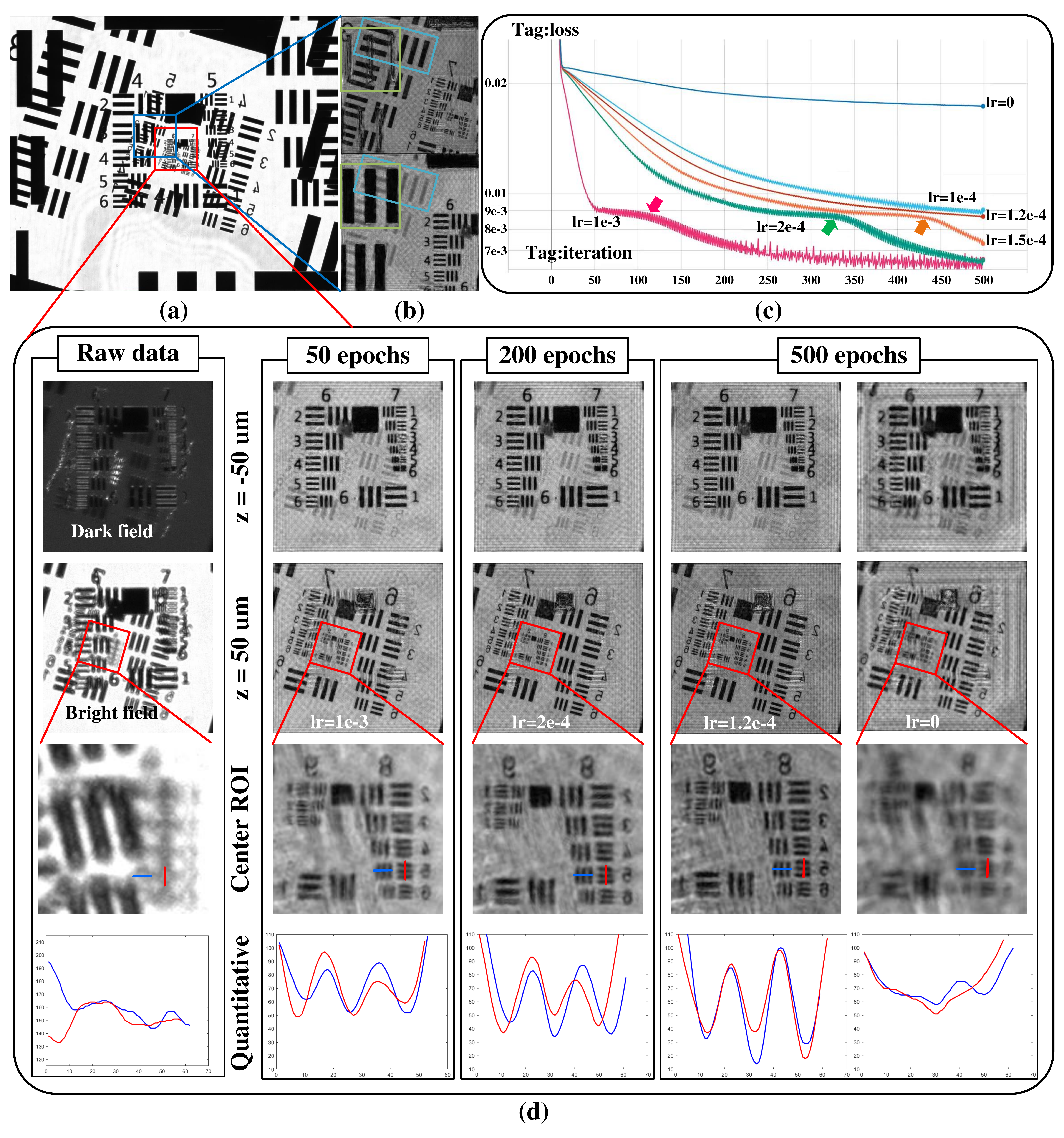}
	\caption{Absorption imaging of a two-slice sample consisting of two resolution targets placed at different depths (-50 and 50 um), with one rotated laterally with respect to the other. (a)Low-resolution full FoV image from 4$\times$ 0.1NA objective. (b)The imaging  for 5-3(12.41um) and 4-3(24.80um) in two slice. (c)The loss curve for different learning rates of light calibration. The overfittings during the optimization are labeled, and the training should be stopped before the overfitting. (d)The captured raw data and the reconstructed results with different learning rates of LICC}
	\label{Fig.4}
\end{figure}

The low-resolution raw images captured consist of 9 brightfield and 216 darkfield images. Since the USAFchart serves as an intensity modulation sample, we employ a complex RI matrix to represent the object. The imaginary parts of the complex RI sections, representing the absorption of the light field at different depths within the sample, are shown in Fig.\ref{Fig.4}(d). MSNN successfully isolates the intensity modulation information at various depths from the raw data.

However, without the LICC, MSNN fails to achieve the theoretical lateral resolution even after 500 optimization iterations, and the imaging quality across depth sections is unsatisfactory. The compromised quality also suggests the presence of system deviations, such as pupil aberration, illumination deviation, and exposure time differences\textsuperscript{\cite{zheng2022robust, zheng2022robust_2}}.

When using the LICC, we manage to reconstruct the sample at an enhanced resolution in as few as 50 iterations. With extended optimization, the lateral resolution improves steadily. However, our experiments show that the optimal learning rate for LICC varies across training iterations. As shown in Fig.\ref{Fig.4}(c), sudden drops in loss occur at marked positions within different loss curves, resulting in a degradation of imaging. Therefore, it's necessary to halt optimization early by observing the loss trend. The early stop point loss within the different curves further supports the notion that LICC improves imaging quality.

\subsection{Analysis of resolution}
The LED array provides various illumination angles to improve the lateral resolution\textsuperscript{\cite{zheng2021concept, zhang2021fast}}. As expected, the lateral bandwidth $\Delta fx$ is determined by the sum of objective $NA(NA_{obj})$ and illuminate $NA(NA_{illu})$:

\begin{align}
\Delta fx = \frac{2(NA_{obj} + NA_{illu})}{\lambda}
\end{align}
The axial resolution, however, does not follow this trend. Its bandwidth, $\Delta fx$, is neither that of the objective nor is it that would result from using an objective having the sum of the two NA. Instead, it is somewhere in between the NA of the objective and the sum of two NAs($NA_{obj} and NA_{illu}$), and can be estimated with a formula:

\begin{gather}
\Delta fz = \frac{2-\sqrt{1-NA^2_{obj}}-\sqrt{1-NA^2_{illu}}}{\lambda}
\end{gather}
Nonetheless, the calculated result using this formula represents the maximum axial resolution achievable across various lateral resolutions. The 3D Fourier diffraction theorem\textsuperscript{\cite{goodman2005introduction}} suggests the lateral resolution and axial resolution in the Fourier Ptychography system for RI tomography are not independent\textsuperscript{\cite{popescu2011quantitative}}. We propose a digital simulation method for more accurate analysis of axial resolution in microscopy systems(details in Supplementary 1). In the USAFchart illuminated by our system, the 5-3 line has 44um theoretical maximum axial resolution, and the 4-3 line pair has 70um theoretical maximum axial resolution. In the green ROI of Fig.\ref{Fig.4}, the information regarding the separation of the 4-3 line pairs at an axial distance of 100um is distinct in different depth sections. In order to further analyze the difference between the axial resolution of our system and the theoretical limit, we narrowed the distance of USAFchart for further experiments(details in Supplementary 1).

\section{Experimental of biological sample}

\subsection{System setup}
The raw data of C.elegan were obtained from Laura Waller's Computational Imaging Lab at UC Berkeley\textsuperscript{\cite{chowdhury2019high}}. Since the diameter of C.elegans usually does not exceed 30um, high-NA microscope objectives are used for high axial resolution.
Consequently, all acquired images are brightfield images, and the raw data were applied HDR combined to calibrate the intensity between different illumination. Therefore, we do not use LICC for adaptive intensity calibration. Simultaneously, we divide the imaging depth of 24um or 30um into 121 layers, where the impact of backscattering becomes more pronounced. So we employ LIAC to estimate the attenuation of the forward propagation field caused by backscattering. The computing platform is aligned with the simulation. The total imaging time to complete 50 iterations of a 1200$\times$1200$\times$121 voxels volume was 3 hours.

\subsection{Experiment results}

\begin{figure}
	\centering
	\includegraphics[scale=0.12]{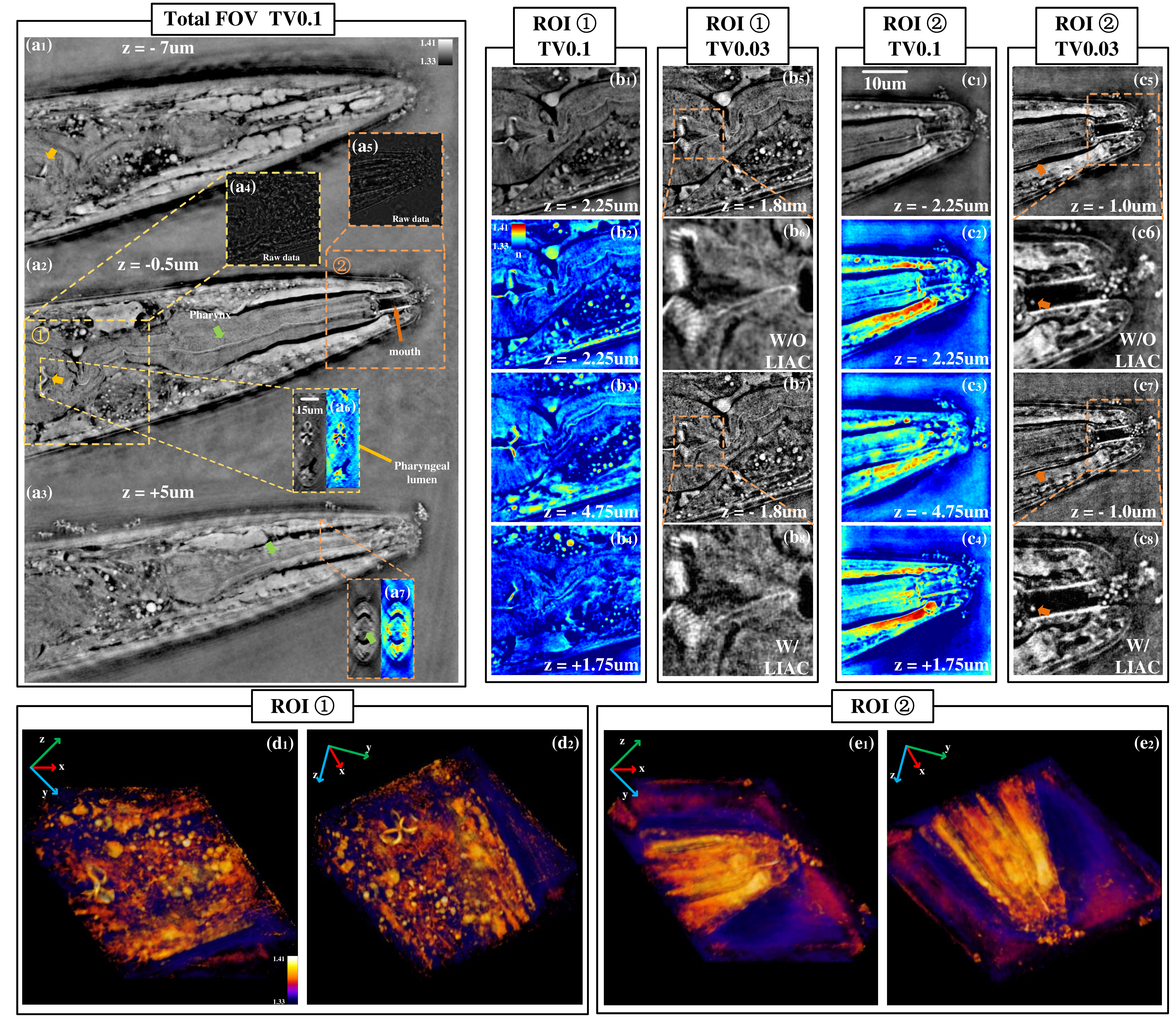}
	\caption{RI tomography of a C.elegan worm's head. (a1)-(a3) Lateral slices through the 3D imaging volume at various axial positions. Major components of the digestive system are labeled. (a4)-(a5)The captured raw data illuminate by the positive incidence. Axial slice through the pharynx (a6), and axial slice through the pharyngeal lumen (a7) are shown. (b1)-(b4),(c1)-(c4) Lateral slices of the ROIs \textcircled{1} and \textcircled{2} in grayscale or RGB color at different axial positions. (b5)-(b8), (c5)-(c8) A comparison of experimental results with and without the implementation of LIAC.(d1)-(d4), (e1)-(e4) 3D reconstructions of the ROIs \textcircled{1} and \textcircled{2} from various viewing angles.}
	\label{Fig.5}
\end{figure}
The captured raw data includes 120 images with illumination angles scanned on a spiral trajectory. Our biological experiment generates a high-resolution RI tomography of an adult C.elegans worm's head region(1200$\times$1200$\times$60 voxels), with voxel size 0.12$\times$0.12$\times$0.5u$m^{3}$.We note that 8-20 iterations are sufficient for visualizing the C.elegan's 3D tomographic structure, taking 25-70 minutes(Visualization 1). MSNN can balance the time required with the quality of 3D imaging. Additionally, as we can utilize various propagation models by substituting the connections within the layers in MSNN, we also conducted experiments for applying the multi-layer Born model\textsuperscript{\cite{chen2020multi}} in MSNN framework(Supplementary 2).

To quantitatively visualize the worm’s 3D biological RI, we present RBG-colored cross-sectional images of all ROIs at various axial and lateral positions. In Fig.\ref{Fig.5}(a1)-(a3), we show two lateral slices through reconstruction volume at the axial position of z=-7, -0.5, +5um. The digestive system including the mouth, pharynx, and pharyngeal lumen, are identified. Behind the pharynx, we can observe the nematode's intestinal lumen, which contains many microsphere-like details.
Fig.\ref{Fig.5}(a6) and (a7) show the axial cross-section of the pharyngeal lumen and pharynx, respectively. In Fig.\ref{Fig.5}(b1)-(b4), the lateral slice clearly show the pharyngeal lumen, pharynx ,and other micron-sized structure at different axial position. We can also see the phenomenon of C.elegan feeding in Fig.\ref{Fig.5}(c1)-(c4). The microspheres in front of the mouth of C.elegan are E.coli bacteria, a food source for the worm(3D RI Tomography results in Visualization 2).

Since the 3D RI of samples are separated into many slices, even for weakly scattered samples, backscattering at such long axial distances is still not negligible, so we conducted comparative experiments with and without LIAC. The results are shown in Fig.\ref{Fig.5}(b5)-(b8) and (c5)-(c8). As observed, the results with LIAC demonstrate enhanced clarity and higher contrast. During optimization, the implementation of LIAC also resulted in a smaller L1 loss compared to raw data(Supplementary 2).

Fig.\ref{Fig.5}(d1)-(d4) and Fig.\ref{Fig.5}(e1)-(e4) show 3D visualizations from various viewing angles. To emphasize the 3D structure, we have obscured parts with an RI of less than 1.33. Usually, visualizing this morphology requires cross-sectional slicing and viewing by electron microscopy. We demonstrate that we can achieve non-invasive imaging to obtain 3D microscopic structures(Visualization 3).

\section{Discussion and Conclusion}
An untrained physical based network termd as MSNN  is proposed and demonstrated in this paper. It is an implementation of the paradigm of remodeling neural networks with optical principles. The scalability of MSNN provides a way to solve the constraints for RI tomography.

To enhance the clarity of RI tomography, darkfield images information is also very important. The implement of LICC can effectively calibrate the intensity between brightfield and darkfield, further, LICC can calibrate the intensity between different positions when using various LEDs(light sources) as illumination, the application of LICC has the ability to reduce hardware requirements and the need for preprocessing of raw data in RI tomography.

Furthermore, due to the complexity of multiple scattering, RI tomography is often limited to imaging samples with weak backscatter under the 1st Born or Rytov approximation. In the future, replacing the backscatter intensity estimation LIAC with a more accurate optical model may enable MSNN to account for higher-order scattering processes, showing promise in reconstructing samples with a broader RI range or extended axial distances.

In this paper, we have introduced MSNN, a neural network for efficient and accurate RI tomography. Compared to the widely used conventional deep learning methods, MSNN reconfigures the conventional CNN with the physical beam propagation model to confirm the authenticity of the 3D RI information. Moreover, it accelerates RI tomography compared to CNNs that employ physical method verification. MSNN introduces a novel paradigm for RI tomography, enabling to image biological samples(C.elegans) that are both thicker and exhibit higher scattering properties than were achievable with earlier methods. The implementation of LICC and LIAC also improved the robustness and imaging clarity and present its scalability. We provide link to datasets and an open-source implementation of MSNN at GitHub repository available at https://github.com/yang980130/Physics-based-3D-tomography-Multi-slice-neural-network. See also the Supplementary Material for supporting content.

\clearpage
\section{Refractive index tomography with a physics based optical neural network: supplemental document 1}
\begin{abstract}
This document provides supplementary information to "Refractive index tomography with a physics based optical neural network".  It first elucidates the concept of the Ewald sphere used in analyzing axial resolution. Following this, it establishes the axial resolutions corresponding to various lateral resolutions, based on the parameters derived from the actual system. 
To validate the congruence between the experiment of multi-slice neural network(MSNN) and the theoretical in axial resolution values, we used an microscope fitted with a 4x objective. We conducted a reconstruction of distinct depth slices, which were formed by a stacking of two USAF charts with a 50-micrometer gap.
\end{abstract}
\maketitle

\subsection{The formulation of Ewald sphere}
The three-dimensional Fourier diffraction theorem suggests the lateral resolution and axial resolution in the intensity diffractive tomography(IDT) system are not independent, but the structures in the biomedical sample with large lateral size usually have a large axial size, we usually ignored studying the relationship between the axial resolution and lateral resolution in our experiment system.

To analyze both lateral resolution and axial resolution theoretically, we use the three dimensional(3D) coherent transfer function(CTF) of the imaging system. As illustrated in Fig.\ref{Fig.1}(a), the 3D spatial frequency(k-space) is formed by the 3D Fourier transform of the object's refractive index(RI)\textsuperscript{\cite{Ewald1969introduction, cowley1995diffraction}}. This spherical k-space representation, known as the Ewald sphere, has a radius determined by the wave vector $k_{0}=\frac{2\pi}{\lambda}$.

By sequentially activating each light-emitting-diode(LED) at different positions on the LED array, which in turn provides plane waves $k_{i}$ at varying angles, we can explore different regions of the k-space. As $k_{i}$ changes with the illumination angle, the maximum probed k-space falls within a spherical shell with radius $k_{0}$. The center of this shell shifts along a second spherical shell(of the same radius $k_{0}$) determined by the incident angle of the plane wave $k_{i}$(as depicted by the gray circle in Fig.\ref{Fig.1}(a)).

However, in actual experiment systems, The Fourier diffraction theorem suggests that, limited by illumination and microscopy, only partial spherical cap bounded by the generalized aperture can be probed. Illuminating the object at different angles will shift different regions of the object’s k-space into a fixed microscope objective lens with fixed numerical aperture(NA). Ultimately, only a portion of the Ewald sphere can be reconstructed.\textsuperscript{\cite{zuo2020wide,horstmeyer2016diffraction}}.

We employ matrices to digitally simulate the 3D CTF based on our system setup. The 3D k-space bandwidth in the x-z section(which is equivalent to the bandwidth in the y-z section) is depicted in Fig.\ref{Fig.1}(b). From this, we can compute the theoretical axial resolution by assessing the k-space bandwidth of the x-z region. The results suggest that different lateral resolutions correspond to varying axial resolutions. This implies that the imaging quality for line pairs on a USAF chart will differ across the depth(z) section due to these resolution disparities.

\begin{figure}
	\centering
	\includegraphics[scale=0.35]{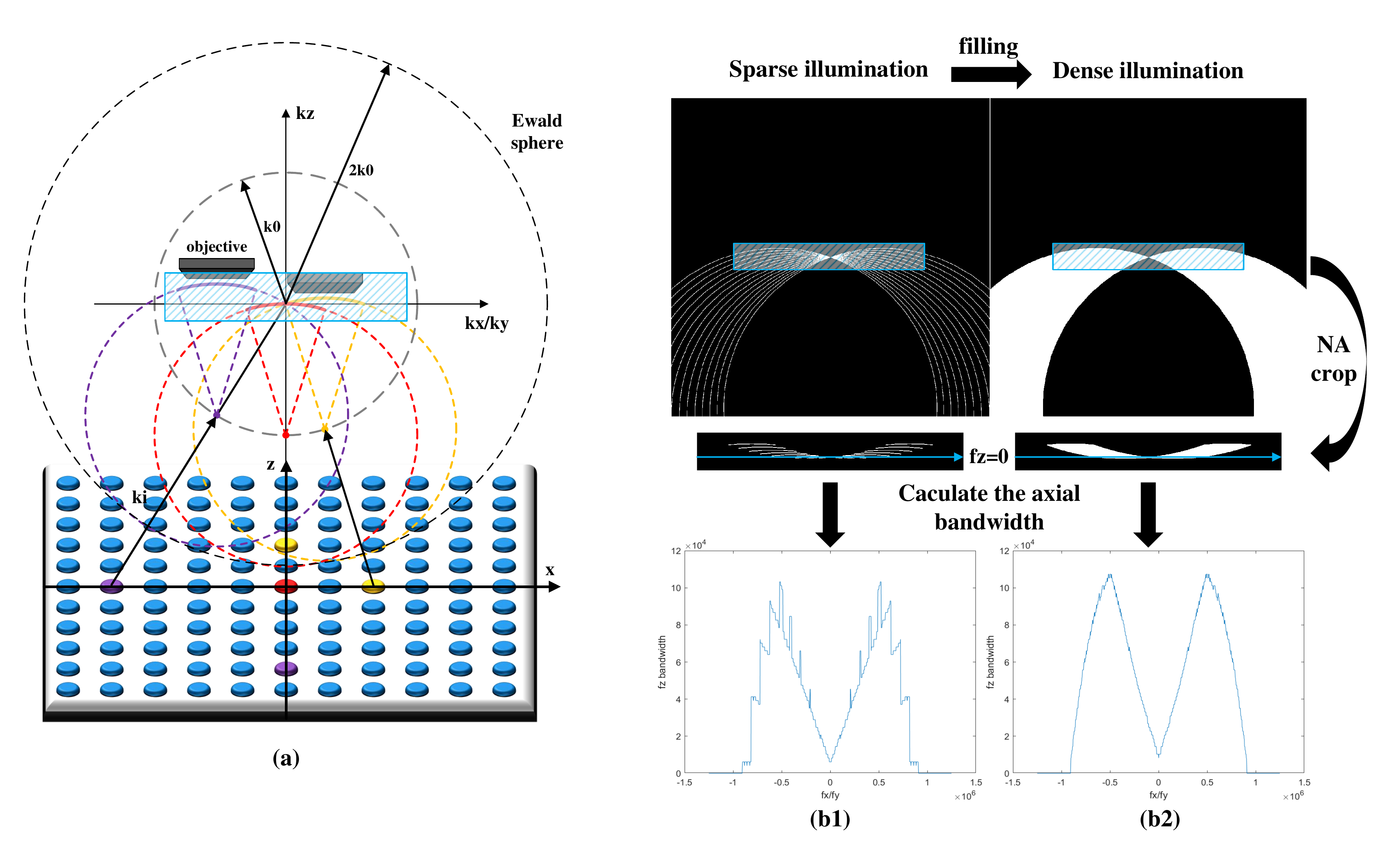}
	\caption{Fourier diffraction theorem in finite-aperture optical systems. (a)The 3D sample is illuminated by plane waves from various angles. The forward scattering wave captured by the aperture of the objective with fixed NA. (b1)The bandpass on the Ewald sphere and the axial bandwidth in different lateral resolutions by the sparse illumination of (a). (b2)The bandpass on the Ewald sphere and the axial bandwidth in different lateral resolutions by the dense illumination}
	\label{Fig.1}
\end{figure}

\begin{figure}
	\centering
	\includegraphics[scale=0.108]{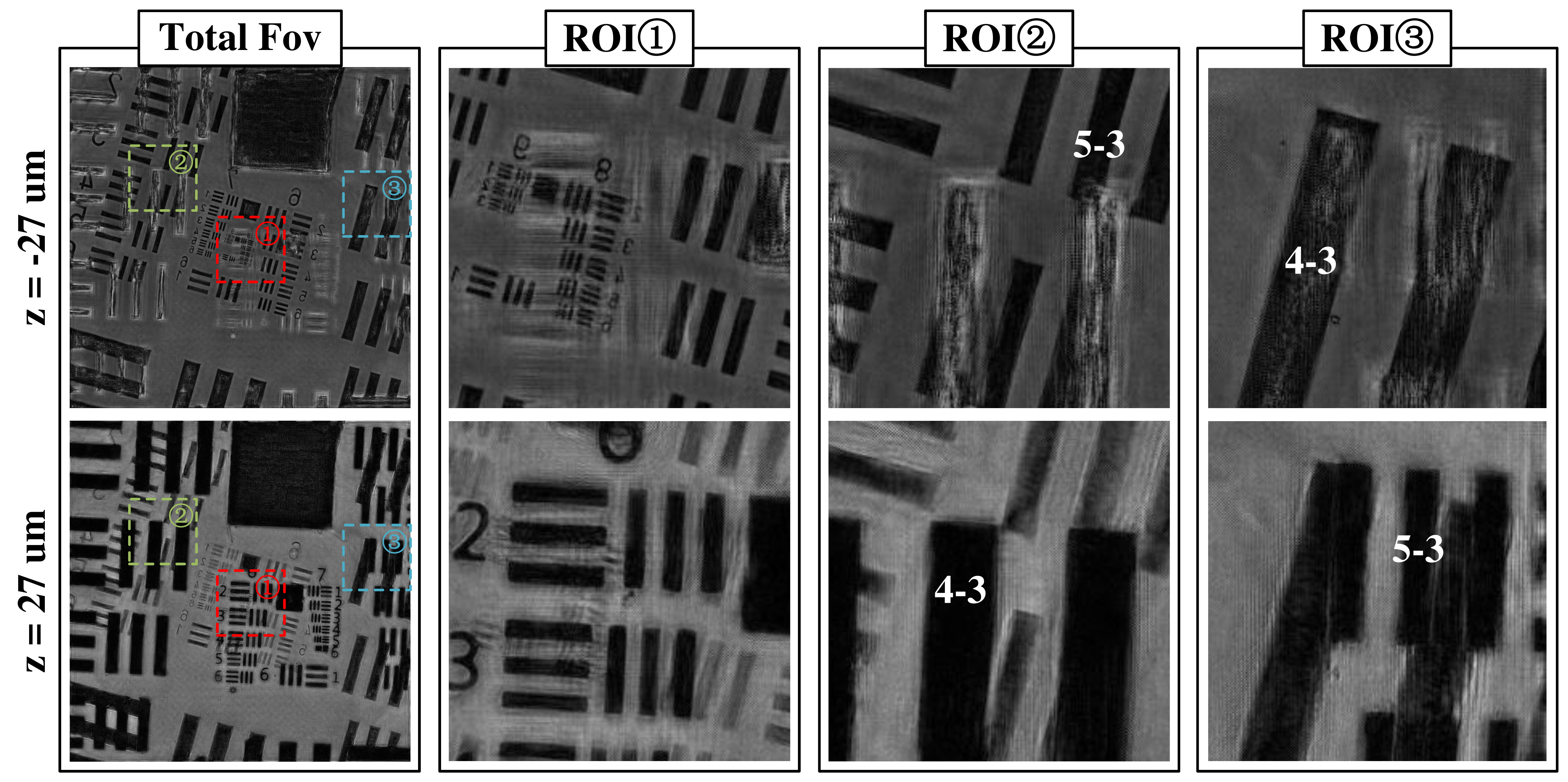}
	\caption{Absorption reconstruction of a two-slice sample consisting of two resolution
		targets placed at different depths(-27 and 27 um), with one rotated laterally with
		respect to the other.}
	\label{Fig.2}
\end{figure}

\subsection{Analysis of axial resolution for the system of quantitative experiment}
Since the lines with the different lateral sizes in USAFchart have the same axial size, the stacked resolution targets provide a convenient way to experimentally characterize lateral resolution at multiple depths. At the same time, we can transform the various linewidth which represents various lateral resolutions into the spectrum and view the reconstruction of depth(z) section to analyze the axial resolutions.

According to the axial bandwidth in Fig.\ref{Fig.1}(b1) which uses sparse illumination, the maximum axial resolution is 9.6um in 1.9um lateral resolution(8-1 line pair). The 5-3 line pair with 12.41um in Fig.\ref{Fig.1}(b) has 57um axial resolution, the 4-3 line pair with 24.8um in Fig.\ref{Fig.1}(b) has 97um axial resolution. In dense illumination, the maximum axial resolution is 9.3um in 2um lateral resolution. The 5-3 line has 44um axial resolution, and the 4-3 line pair has 70um axial resolution. The difference between axial resolutions in low-resolution lines is mainly caused by numerical simulation errors(The numerical error can be reduced by expanding the matrix scale), but the analysis still can be used as a reference. The actual axial resolution should lie between the above two sets of data, taking into account the density of the illumination.

To confirm whether the axial resolution of a 4-3 or 5-3 pair aligns with the theoretical analysis value, we present the reconstruction results of a pair of USAF charts, reconstructed by MSNN, with a placement spacing of 54 um. As observed in Fig.\ref{Fig.2}, ROI\textcircled{1} with smaller lateral sizes demonstrate better layering compared to ROI\textcircled{2} and ROI\textcircled{3}, which have larger lateral sizes. And due to the axial bandwidth not being symmetrical, ROI\textcircled{2} and ROI\textcircled{3}, which have the same lateral size, also exhibit different layering effects in planes at varying depths.

\clearpage
\section{Refractive index tomography with a physics based optical neural network: supplemental document 2}

\begin{abstract}
This document provides supplementary information to "Refractive index tomography with a physics based optical neural network". In this supplementary material, we elaborate on how to approximate backscattering between layers following the scattering principle. We present experimental comparisons of results produced by the multi-slice neural network(MSNN) with and without the use of adaptively adjusted backscatter(LIAC), specifically in scenarios without the employment of Total Variation(TV) regularization. 
Furthermore, we conducted the application of the multi-layer Born model within the MSNN framework. This supplementary concludes with a comparison between the imaging of the multi-layer Born model and beam propagation model both applied to the MSNN.
\end{abstract}
\maketitle

\subsection{The formulation of Ewald sphere}
In 3D RI imaging, modeling the multi-scattering of light field in the sample is highly challenging. The discrepancy between the model and the actual physical process will manifest as noise in the 3D RI results. So researchers usually use the scattering model with the 1st Born approximation or employ the beam propagation model \textsuperscript{\cite{maiden2012ptychographic, chowdhury2019high}} for the prediction of the field in weakly scattered samples. However, even with these methods, only a partial prediction of the scattering field is possible. In the MSNN, we divide the 3D RI of samples into slices, assuming that each slice is thin enough to disregard the scattering within it. The scattering model suggests that the phases of the forward and backscattering fields are identical\textsuperscript{\cite{chen2020multi}} when the slices are extremely thin:

\begin{figure}
	\centering
	\includegraphics[scale=0.39]{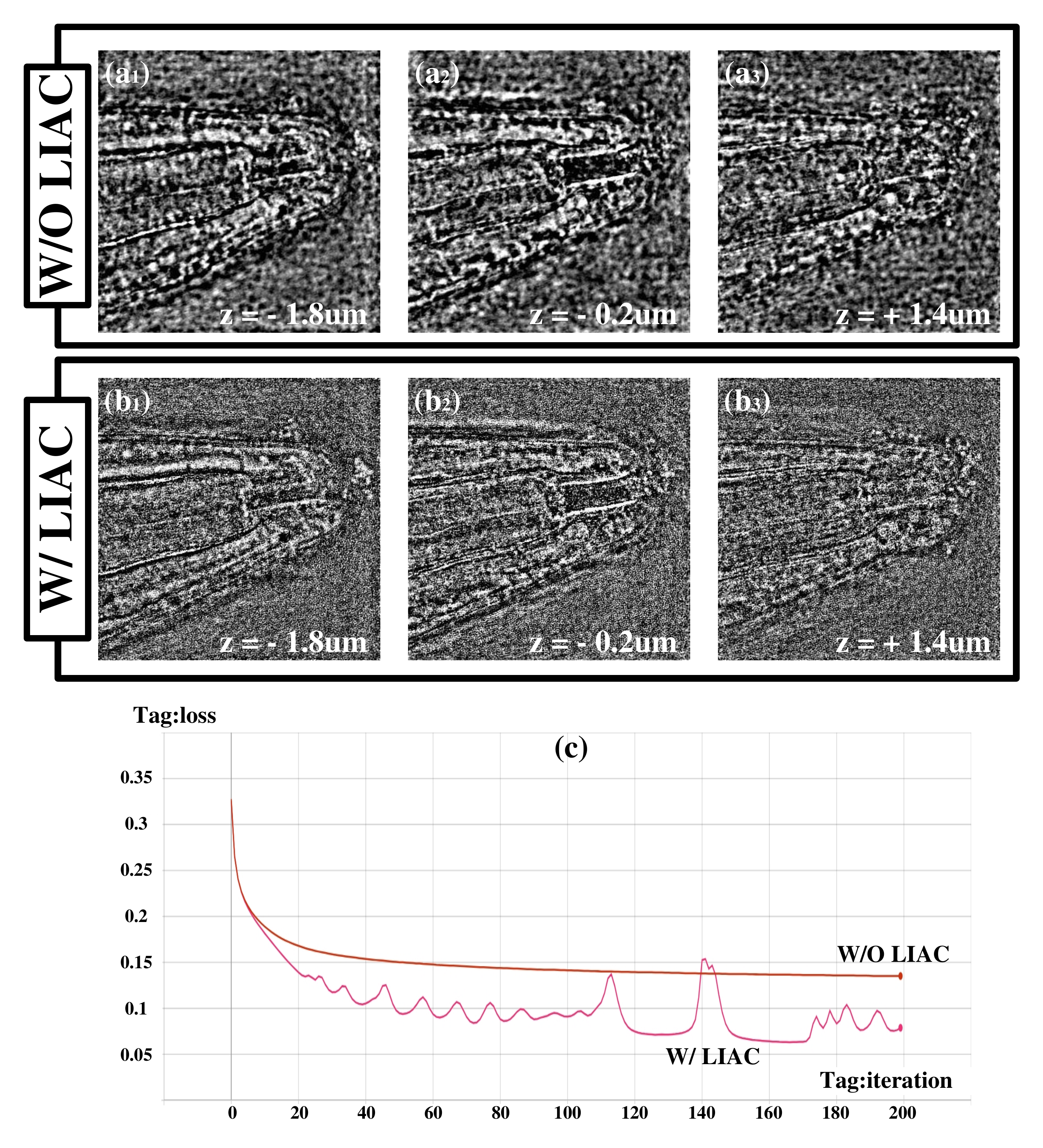}
	\caption{Imaging results without TV regularization.(a1)-(a3)Lateral slices through the 3D reconstruction volume without the implementation of LIAC at various axial positions.(b1)-(b3)Lateral slices through the 3D reconstruction volume with the implementation of LIAC at various axial positions.(c)Loss curve of the optimizations}
	\label{Fig.1}
\end{figure}

\begin{figure}
	\centering
	\includegraphics[scale=0.21]{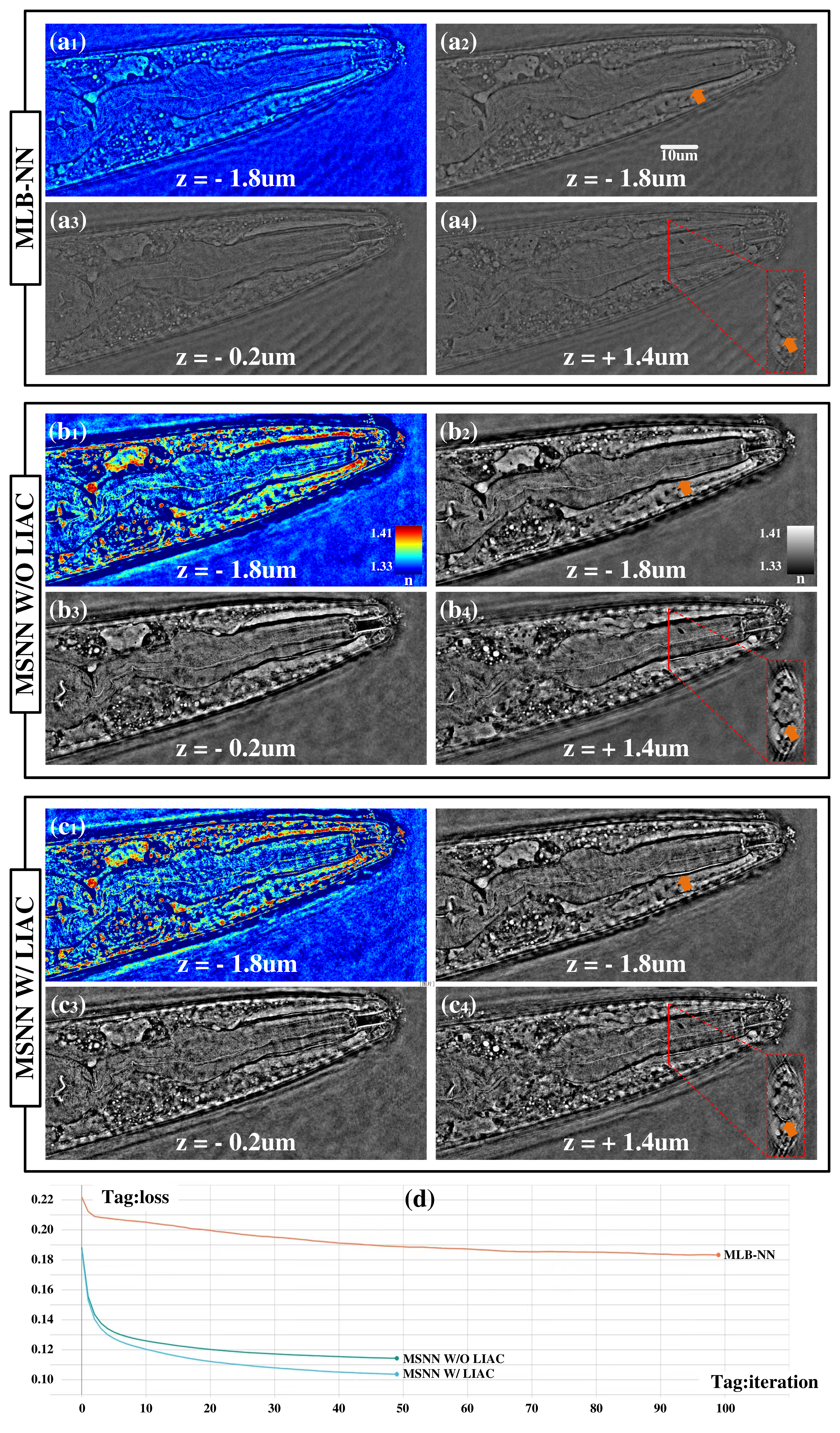}
	\caption{The comparison of experimental results of a C.elegan worm's head with different methods. (a1)-(a4)Lateral slices through the sample imaging by neural network using multi-layer Born propagation model(MLB-NN). (b1)-(b4)Lateral slices through the sample imaging by the MSNN with LIAC. (c1-c4)Lateral slices through the sample imaging by the MSNN without LIAC. (d)Loss curves of the optimizations}
	\label{Fig.2}
\end{figure}
\begin{gather}
U^{(born)}_{s}(r, z) = \tilde{G}(u, \Delta z) \iint sinc(\daleth(u')+\daleth(u))\Delta z) \tilde{U}_n(u') \tilde{V}_n(u-u') \Delta z d^2u' \\
U^{(born)}_{bs}(r, z) = \tilde{G}(u, \Delta z) \iint sinc(\daleth(u')-\daleth(u))\Delta z) \tilde{U}_n(u') \tilde{V}_n(u-u') \Delta z d^2u' \\
\daleth(u) = (\sqrt{\dfrac{n_m}{\lambda}-||u||^2})
\end{gather}
where $\Delta z$ is the thickness of the slice, allowing us to approximate the $sinc$ function to one. By applying this conclusion to the beam propagation model, we can consider that the backscattering field attenuates the forward scattering field, although the intensity of this attenuation remains unknown. To make the beam propagation model more closely resemble the actual propagation of the light field, we employ LIAC to predict the intensity of the backscattering field and attenuate the forward scattering fields, thereby enhancing the RI image quality. To analyze the impact of LIAC on the RI image quality for samples, we used MSNN to optimize 200 iterations without applying TV regularization. The total imaging time to complete 200 iterations of a 420$\times$420$\times$60 voxels volume(The mouth of C.elegan) was 44 minutes. And the imaging results are shown in Fig.\ref{Fig.1}.

As observed, the loss curves indicate that the imaging results incorporating LIAC for predicted intensity align more closely with the captured raw data in the majority of iterations.
Furthermore, the LIAC-enhanced results exhibit high-frequency noise which can be eliminated by many effective method, and we think that the high-frequency noise is caused by higher-order scattering fields. Imaging results without LIAC show more speckle-like noise, which would be difficult to remove, and we think that this noise is caused by not considering low-order scattering such as backscattering\textsuperscript{\cite{azimi1983distortion, belkebir2006influence}}.

\subsection{The comparison of total field-of-view imaging with total variation regularization}
Considering the challenge of separating information and noise in the reconstructed image without regularization, we present the total field-of-view reconstructed image using TV regularization with or without LIAC. Furthermore, the MSNN allows for the easy exchange of field-propagation models by defining the connections between layers. We also employ the multi-layer Born model within the Deep Neural Network(MLB-NN) for imaging the sample\textsuperscript{\cite{chen2020multi}}, as described in the methodology chapter of the text.

As shown in Fig.\ref{Fig.2}, the imaging utilizing MLB-NN shows a similar RI distribution to MSNN, albeit with lower contrast, and the loss converges at a slower pace. With the same level of TV regularization applied, the reconstructed image using MSNN with LIAC exhibits a quicker convergence speed and superior image granularity in the lateral slices and axial slices, which may reveal more details(labeled in Fig.\ref{Fig.2}) in the tomography.

\bibliographystyle{unsrt}
\bibliography{references}
\end{document}